\documentclass[11pt]{article}

\newcommand{\be}{\begin{eqnarray}}
\newcommand{\ee}{\end{eqnarray}}
\newcommand{\rar}{\rightarrow}

\begin{document}

\begin{center}
{\bf \large{Screening of magnetic fields by charged Bose condensate.
}} \\ \vspace{0.5cm}
{\it Alexander D. Dolgov \footnote{dolgov@fe.infn.it}$^{a,b,c}$
and Angela Lepidi \footnote{Angela.Lepidi@physik.uni-muenchen.de}$^{d}$}
\\  \vspace{0.3cm}
$^a$ Istituto Nazionale di Fisica Nucleare, Sezione di Ferrara, 
I-44100 Ferrara, Italy \\
$^b$ Dipartimento di Fisica, Universit\`a degli Studi di Ferrara, 
I-44100 Ferrara, Italy \\
$^c$ Institute of Theoretical and Experimental Physics, 113259 Moscow, Russia \\
$^d$ Arnold-Sommerfeld-Center for Theoretical Physics, Department
f\"ur Physik, Ludwig-Maximilians-Universit\"at M\"unchen, Theresienstr. 37,
D-80333, Munich, Germany

\begin{abstract}
The space-space component of the photon polarization operator is 
calculated in zero frequency limit  for a medium with Bose-Einstein condensate (BEC) of electrically charged 
particles. It is found that the polarization operator tends to a finite value at vanishing photon 
3-momentum, as it happens in superconducting media. It means that magnetic fields are 
exponentially screened in such a medium analogously to the Debye screening of electric 
charges. At non-zero temperature the  screened  magnetic field oscillates and contains a
contribution which drops only as a power of distance. This phenomenon is unknown for
superconductors, even in BEC phase and can be potentially observable.
\end{abstract}
\end{center}


As is well known, electrically charged impurities in plasma are screened according to the Debye law~\cite{debye}:
\begin{eqnarray}
	\label{yukawa}
	U = \frac{q}{4\pi}\,\frac{\exp(-m_D r)}{r}\,,
	\end{eqnarray}
where $q$ is the charge of a test particle and $m_D$ is the screening (Debye) mass, which is expressed through the density of charge carriers in plasma. On the other hand, it is known that magnetic fields are not screened in plasma because of the absence of magnetic monopoles. The last statement has been proven in standard QED~\cite{fradkin} and in scalar QED (SQED), when bosons are not condensed.
We show below that, in the presence of Bose-Einstein condensate (BEC) of charged particles, magnetic fields in plasma are exponentially screened. 
Such screening is a manifestation of the well-known Meissner effect in superconductors. However, at non-zero temperature new phenomena of an oscillating exponential screening and a power law one are found here. The screening mass is of course different from the usual Debye one, which, in the relativistic case, is proportional to the temperature or the chemical potential. 
The charged condensate contributes to the photon polarization tensor by additive terms, whose amplitude is proportional to the amplitude of the condensate and is the same for electric and magnetic photons.
The calculations are done in the framework of SQED with charged scalar boson condensate but it is evident that similar screening exists in the presence of charged vector condensate.
Moreover, the result seems to be applicable to non-Abelian gauge theory as well. These problems will be studied elsewhere. 

The electric screening in plasma in the presence of charged Bose condensate was first studied in refs.~\cite{Hore:1975} and recently in relativistic theory in refs.~\cite{Gabadadze:2007si,Dolgov:2008pe,Gabadadze:2008pj,Dolgov:2009yt}, see also \cite{Gabadadze:2009jb} and references therein. It has been discovered that the photon polarization operator in plasma is singular at zero photon 3-momentum, having poles proportional to $1/k^2$~\cite{Dolgov:2008pe,Gabadadze:2008pj} and $1/k$~\cite{Dolgov:2008pe,Dolgov:2009yt}, which lead to an oscillating and power law screening behavior. 
Such screening behavior is in good agreement with the earlier papers~\cite{Hore:1975}. Magnetic, as well as electric, screening with charged BEC at zero temperature was considered also in ref.~\cite{Alexandrov:95}, where the superconducting behavior as a response to electromagnetic field was observed.

The Fourier transform of the Maxwell equations in plasma is modified as: 
\begin{eqnarray}
	\label{Phot_EOM_Pi_munu}
	\left[ k^\rho k_\rho g^{\mu\nu} - k^\mu k^\nu + \Pi^{\mu\nu} (k)\right] A_\nu (k) 
	= \mathcal{J}^\mu (k)\, ,
	\end{eqnarray}
where $A_\mu$ is the electromagnetic potential, $k_\mu$ is the photon 4-momentum, 
$\mathcal{J}^\mu (k)$ is the 4-vector of an electric current, and $\Pi_{\mu\nu}$ is the polarization
operator of photon in medium. The latter can be expressed through the distribution functions
of charged particles and antiparticles in medium, $f(q)$ and $\bar f (q)$, where $q$ is the
particle 3-momentum (see below). 

The expression for the polarization operator can be found in many textbooks on field theory
at non-zero temperature. 
For details of calculations we address to our work~\cite{Dolgov:2008pe}, where the
notation is the same as here, to refs.~\cite{Dolgov:2009yt} and~\cite{Kapusta:1989tk} and
references therein.
In the lowest order in electric coupling, the photon polarization
operator induced by interactions with charged bosons has the form:
\begin{eqnarray}
	\label{phot_pol_tensor_bos}
	\Pi_{\mu\nu}^{B} (k) 
	= e^2 \hspace{-0.1cm} \int \frac{d^3q}{(2 \pi)^3 E} 
	\left( f_B + \bar f_B \right)
	\left[  \frac{1}{2} \, \frac{(2q - k)_\mu (2q - k)_\nu}{(q - k)^2 - m_B^2}
	+\frac{1}{2} \, \frac{(2q + k)_\mu (2q + k)_\nu}{(q + k)^2 - m_B^2}  -   g_{\mu\nu}\right],
	\end{eqnarray}
where index $B$ denotes bosons, $E=\sqrt{q^2+m_B^2}$
and we omitted for brevity the arguments of $f$ and $\bar f$.

In thermal equilibrium the distribution functions
of bosons and anti-bosons, respectively $f_B$ and $\bar f_B$, take the form:
\begin{eqnarray}
 \label{f-B-cond}
	f_B^{(C)} &=& \frac{1}{\exp [ (E-m_B) / T ] - 1} + C \, \delta({\bf q}),\\
\label{bar-f-B}
\bar f_B  &=& \frac{1}{\exp[ (E+m_B) / T ]  - 1},
\end{eqnarray}
where the chemical potential of bosons is  equal to their mass: $\mu = m_B$ 
and $\bar \mu = - m_B$ for anti-bosons. The constant parameter $C$ describes the 
amplitude of the condensate. Note that bosons condense when their chemical potential
reaches the maximum allowed value equal to the boson mass. 

The time-time component of the polarization operator,  $\Pi_{00}$, was analyzed in detail in refs. \cite{Dolgov:2008pe, Dolgov:2009yt}. 
Such a quantity, in the limit of vanishing photon energy, $\omega = 0$, and small magnitude of the photon three-momentum, $k\rightarrow 0$,
is relevant to the screening of electric charge.
The screening of magnetic fields is connected with the space-space components,
$\Pi_{ij}$, which, in the homogeneous and isotropic case, can be presented as:
\begin{eqnarray}
\Pi_{ij} = a(k)\left(\delta_{ij} - \frac{k_i k_j}{{\bf k}^2}\right)
+ b(k) \frac{k_i k_j}{{\bf k}^2}\,.
\label{Pi-decomp}
\end{eqnarray}
To calculate $a(k)$ we first multiply $\Pi_{ij}$ by $\delta_{ij}$ and, second, by $k_i k_j$. Correspondingly we obtain:
\be
2 a(k) +b(k) &=& \frac{e^2}{16\pi^3}\,\int \frac{d^3 q}{E}\,\left(f+\bar f\right)\,
\left[ \frac{4 q^2 -k^2}{2 {\bf k q} - k^2 } -  \frac{4 q^2 -k^2}{2 {\bf k q} + k^2 } +2
\right]\,,
\label{2ab}\\
b(k) &=&0 \,,
\label{b}
\ee
where ${\bf k q} = k q \cos \theta$. Hence
\be
a(k) = \frac{e^2}{32\pi^3}\,\int \frac{d^3 q}{E}\,\left(f+\bar f\right)\,
\left[ 2 +\frac{2 k^2 (4 q^2 -k^2)}{4 ({\bf kq})^2 - k^4 } \right]\,.
\label{a}
\ee

If only the condensate term (delta-function) is retained in distribution function (\ref{f-B-cond}),
the integration over $d^3 q $ is trivial and we obtain: 
\be
a^{(C)} = \frac{ e^2 C} {8 \pi^3 m_B}  \equiv e^2 m_C^2\,.
\label{a-cond}
\ee
Since $a(k)$ does not vanish at $k=0$ but tends to a constant, the asymptotic behavior of 
magnetic fields at large distances from the source would be given by the exponentially
decreasing Yukawa potential, i.e. essentially the same as the Debye screening. 
The existence of the magnetic mass in a medium with a BEC was found in ref.~\cite{Gabadadze:2008pj}, 
but the screening effects were not emphasized there. In the subsequent paper
by the same authors~\cite{Gabadadze:2009qe} it was shown that below certain critical value magnetic 
field is expelled from the condensate, while for larger values it penetrates the condensate within 
flux-tubes.

On the other hand, if we consider the standard Bose distribution with $\mu<m_B$, 
when no condensate is formed, the distribution has the usual form:
\be
f_B = \frac{1}{\exp[ (E-\mu) / T ]  - 1}\,,
\label{f-b}
\ee
then $\Pi_{ij} (k)$ vanishes as $k^2$ in the limit $k\rar 0$, as expected. 
Indeed, the integrals over angles is easily taken and we find:
\be 
a(k) =  \frac{e^2}{16\pi^2}\,\int \frac{q^2 d q}{E}
\,\left(f+\bar f\right)\,
\left[ 4 - \frac{4 q^2 - k^2}{k q}\,\ln \bigg|\frac{2q +k}{2q -k}\bigg| \right]\,.
\label{a-integ}
\ee
Expanding the integrand in in powers of  $(k/q)$ we find: 
\be
a(k) \approx \frac{e^2 k^2}{24 \pi^2}\, \int \frac{dq}{E} \left( f + \bar f \right) \,.
\label{a-no-C}
\ee
As expected, $a(k) \sim k^2$ and magnetic fields are not screened. 
 
Note that expression (\ref{a-no-C}) is singular in the limit $m_B = 0$, since the integral diverges as $1/q^2$ at the lower
bound of integration, $q=0$. 
Even if there are no massless charged particles, somewhat
similar singularity exists for massive ones if $\mu=m_B$ and $k\rar 0$, as it was 
found in ref.~\cite{Dolgov:2008pe}. 
Such singularity leads to a term $\sim 1/k$ in the time-time 
component of the polarization tensor. 
Similarly, the same singularity leads to a non-standard contribution to the space-space component: $a(k) \sim k$, when $k\rar 0$. 
This can be demonstrated by evaluating
the integral in (\ref{a-integ}) with the 
distribution function (\ref{f-b}) where $\mu=m_B$. The additional, harder in $k$, terms
come from the integration region where $q$ is small, i.e. comparable to $k$.  It is convenient
to separate the integral into two parts $0<q<k/2$ and $k/2 <q< \infty$. 
In the first part we introduce  the new integration variable 
$x=2q/k$, so $0<x<1$.  In the limit of small $k$ the energy can be expanded as
$E_B \approx m_B + k^2x^2/8m_B$. At small $q$ the distribution function is infrared singular:
\begin{eqnarray}
\left[\exp\left(\frac{E_B-m_B}{T}\right) -1\right]^{-1} \approx \frac{2m_B T}{ q^2}= 
\frac{8m_B T}{k^2 x^2}\,.
\label{exp-expand}
\end{eqnarray}
Usually this singularity is not dangerous because it is canceled by the integration 
measure, $\sim q^2$. However, the logarithmic term behaves as
$k/q$ for $q>k$ and as $q/k$ for $q<k$. Thus the integral is finite, but it does not vanish
as $k^2$ when $k\rar 0$.

The first part of the integral with $q<k/2$ can be taken analytically and we obtain: 
\begin{eqnarray}
a^{(s)}_1 (k) = \frac{e^2 kT}{8\pi^2}\,\int_0^1 dx \left[ 2 - \left(x -\frac{1}{x}\right) 
\ln \bigg| \frac{1+x}{1-x}\bigg| \right] =   \frac{e^2 kT}{
8\pi^2}\,\left(1+\frac{\pi^2}{4}\right)\,.
\label{a1}
\end{eqnarray}
There is also  another contribution coming from the part of the
integral with $q>k/2$. 
As $k\rar 0$, the second part of the integral, $k/2 <q< \infty$, gives: 
	\begin{eqnarray}
	a^{(s)}_2 (k) = \frac{e^2 kT}{8\pi^2}\,\int_1^\infty dx \left[ 2 - \left(x -\frac{1}{x}\right) 
	\ln \bigg| \frac{1+x}{1-x}\bigg| \right] =   \frac{e^2 kT}{8\pi^2}\,\left(-1+\frac{\pi^2}{4}\right)\,,
	\label{a2}
	\end{eqnarray}
such that the total contribution is:
	\begin{eqnarray}
	\label{a-tot}
	a^{(s)}(k) = a^{(s)}_1 (k) + a^{(s)}_2 (k) = \frac{e^2 T}{16}\, k\,.
	\end{eqnarray}
For small $k$ this term could dominate over the usual $k^2$ term and would change the 
screening behavior. 

In the transverse gauge, $k_\mu A^\mu = - k_j A_j =0$, 
equation (\ref{Phot_EOM_Pi_munu}) takes the form: 
\be
\left[ k^2 + a(k) \right] A_i =   \mathcal{J}_i (k)\, ,
\qquad 
a(k) = a^{(s)} (k)+ a^{(C)} (k)
\label{eq-trans-gauge}
\ee
whose solution in coordinate space can be written in terms
of the Green's function as: 
\be
A_i (x) = \int d^3 y\, G(x-y)\, \mathcal{J}_i(y) \,.
\label{green-fun}
\ee
The asymptotic behavior of $G$ 
is determined by the integral (see e.g. ref.~\cite{Dolgov:2009yt}):
\begin{eqnarray}
G (r) =
\frac{(-i)}{4\pi^2 r} \, \int_0^\infty \frac{dk k\,\left(e^{ikr}-e^{-ikr}\right) } {k^2 + a (k)}\,.
\label{U-of-r}
\end{eqnarray}
According to the result obtained above, $a(k)$ may contain odd terms in $k$ and hence
the integral along the half real $k$-axis cannot be extended to the whole real axis, so we 
present the denominator as half of sum and difference of even and odd function as following:
\be
f(k) = [ f(k) + f(-k)]/2+ [f(k) - f(-k)]/2 \,.
\label{f-of-k}
\ee
Since $a(k) = k^2 + e^2 m^2_C + e^2 Tk /16$, eq. (\ref{U-of-r}) can be rewritten as: 
\be
G (r) =
\frac{(-i)}{4\pi^2 r}\, \int_0^\infty \frac{dk k\,\left(e^{ikr}-e^{-ikr}\right) \left(k^2 + e^2 m_C^2 - e^2 T k/16 \right)} 
{(k^2+e^2 m_C^2)^2  - e^4 T^2 k^2 /256 }\,.
\label{G-symm-asym}
\ee
The integral of the even part may be transformed, as usually,  into the integral along  the whole
real axis and after closing the contour in the upper (for $e^{ikr}$ ) or lower (for $e^{-ikr}$ ) half-plane
we express the result through the residues in the corresponding poles in the complex $k$-plane at: 
\be
k^{(pole)} = \pm i \sqrt{e^2 m_C^2 - \frac{ e^4 T^2}{1024}}  \pm \frac{e^2 T}{32}\,.
\label{k-pole}
\ee
If $ m_C > e^2T/32$, the resulting  screened potential would be exponentially cut with 
superimposed oscillations. For $ e^2 T \ll 32\,  m_C $, the Green function takes the form:
	\begin{eqnarray}
	G(r) \sim \exp (-e m_C r)\,\cos (e^2 r T/32)\,.
	\end{eqnarray}
In this case the spatial damping scale is much shorter than the oscillation scale. However, if $e T \sim m_C$
the scales are comparable. These oscillations remind
the Friedel oscillations~\cite{friedel} in fermionic plasma, but physics is different.

The contour of the integration of the second, odd in $k$, part can be closed in the upper or lower quadrant
of the complex $k$-plane. So in addition to the poles in these quadrants we need to include the contributions
from the integrals over the imaginary axis. The latter produces a power law asymptotics of $G(r)$.
If the condensate is absent, but $\mu=m_B$,
then at small $k$: $a(k) \approx e^2 k T /16$, see eq. (\ref{a-tot}).
Correspondingly at large $r$ the Green's function drops as: 
$G(r) \sim  8/(\pi^2e^2 r^2T)$. This asymptotic behavior is realized when $r> 1/T$. 
In the presence of the condensate the Green's function acquires an additional constant term 
$e^2 m_C^2$ (\ref{a-cond}). In this case the contribution of the integral over the imaginary axis of $k$
 gives  $G\sim  T/(16 \, e^2 \pi^2 r^4 m_C^4)$.

In ref.~ \cite{Dolgov:2010gy} it was found that 
$W$ bosons, which may condense in the early universe in the presence of a large lepton asymmetry \cite{Linde:1976kh}, would form a ferromagnetic state with spins aligned on macroscopic scales. 
As it was pointed out in the quoted paper, the ferromagnetic alignment could be destroyed by screening of magnetic interactions.
By comparing the screening length $\lambda \equiv 1/e m_C$ with the typical inter-particle distance $d \equiv C^{-1/3}$, one can roughly check whether screening is effective or not. 
It follows that screening is effective unless $m_B /C^{1/3} \ll 10^{-4}$. 
Thus it seems hardly possible 
 that the ferromagnetic state could be realized in the broken phase of the electro-weak sector, where the boson mass, $m_B$, is essentially determined by the Higgs VEV. 
On the other hand, screening may be ineffective in the unbroken phase, where $m_B$ is much smaller and determined by radiative corrections.

The exponential screening of magnetic fields and their oscillations in massless scalar electrodynamics 
were mentioned in ref.~\cite{kraemer}. They were justly prescribed to an inaccurate treatment of the perturbation 
theory with massless charged bosons. If one takes into account the temperature corrections to the
charged boson mass by the resummation of the perturbation series, the effect disappears and the 
polarization operator vanishes as $k^2$ at small $k$ in accordance with the expectations.
However, in our case the temperature corrections to the mass do not help to avoid the effects 
considered here. According to kinetic treatment,  the condensate appears when the boson chemical
potential, $\mu$, is equal to the boson mass. So the temperature corrections to the mass are canceled by equal correction to 
the maximum value of $\mu$ and the singularity is not eliminated.

The exponential screening in the considered system indicates that the system is similar to the superconductor
in strongly coupled BEC phase. However, the new effects of power  law decrease of the magnetic
field and its oscillations are new and quite surprising. Possibly the same or analogous phenomena exist 
in the usual superconductors and may be observable.

The phenomenon of the power law  magnetic field screening may have interesting applications to 
compact stars with magnetic fields. This effect could also give rise to an infrared cutoff in 
non-Abelian gauge theories, which suffer from very strong infrared singularities~\cite{linde-1979}. 
Moreover, the infrared divergences in non-Abelian theories could enforce or facilitate the
formation of Bose-Einstein condensation of charged bosons.

\vspace{0.5cm}
{\bf \noindent Acknowledgements}
A.L. acknowledges Denis Comelli for insightful discussions. We also thank G. Gabadadze
and B. Shklovsky for helpful comments.
We thank the anonymous referee for informing us about earlier works \cite{Hore:1975,Alexandrov:95}. The work of A. L. is supported by TRR 33 ''The Dark Universe''.


\end{document}